\documentclass[
    ,final            
  ]
  {aipproc}

\layoutstyle{6x9}

\def\hs{\qquad}               
\def\nn{\nonumber}            
\def\beq{\begin{eqnarray}}    
\def\be{\begin{eqnarray}}
\def\eeq{\end{eqnarray}}      
\def\ee{\end{eqnarrayn}}
\def\at{\left(}               
\def\aq{\left[}               
\def\ct{\right)}              
\def\cq{\right]}              

\def\R{{\hbox{{\rm I}\kern-.2em\hbox{\rm R}}}}   
\def\H{{\hbox{{\rm I}\kern-.2em\hbox{\rm H}}}}   
\def\N{{\hbox{{\rm I}\kern-.2em\hbox{\rm N}}}}   
\def\C{{\ \hbox{{\rm I}\kern-.6em\hbox{\bf C}}}} 
\def\Z{{\hbox{{\rm Z}\kern-.4em\hbox{\rm Z}}}}   
\def\dir{/\kern-.7em D\,}                          
\def\lap{\Delta\,}                                 

\def\de{\delta}

\def\Ga{\Gamma}

\def\La{\Lambda}

\def\be{\begin{equation}}
\def\ee{\end{equation}}
\def\bea{\begin{eqnarray}}
\def\eea{\end{eqnarray}}

\def\nn{\nonumber}

\begin{document}     

\title{Stability in Generalized  Modified  Gravity}

\classification{11.25.-w, 95.36.+x,98.80.-k  }
\keywords      {modified gravity, stability}

\author{S. Zerbini}{
  address={Department of Physics, University of Trento, and\\
  Gruppo Collegato INFN di Trento, Sezione di Padova, Italy},
}

\begin{abstract}
The stability issue of a large class of modified gravitational models is 
discussed with  particular emphasis to de Sitter solutions. Three approaches 
are briefly presented and the generalization to more general cases 
is mentioned.

\end{abstract}

\maketitle


\section{Introduction}
It is well known that recent astrophysical data are in agreement with a 
universe in current phase of accelerated expansion,
in contrast with the predictions of Einstein gravity in FRW  space-time.
It seems that the most part of energy contents (roughly $75\%$) in the universe is due to  mysterious entity  with negative pressure: the so dubbed Dark Energy.
The simplest explanation is $\Lambda$CDM: Einstein gravity plus a small 
positive cosmological constant 
suffers from the coincidence problem and  the
cosmological constant issue: $\Lambda_{ob}/\Lambda_{th}\simeq 10^{-120}$,
while if we assume supersymmetry, one has 
 $\Lambda_{ob}/\Lambda_{th}\simeq 10^{-60}$. In this case 
$w_{de}=\frac{p_{de}}{\rho_{de}}=-1$.
There exist alternative  explanations. Among  many, we recall:

{\bf i. Modification of gravity on large scale: DGP brane-world model} 
\cite{d}.

{\bf ii. Dark energy associated with cosmological scalar fields,
 quintessence if  $w_{de}>-1$, phantom matter if  $w_{de}<-1$.}  

{\bf iii. Modified gravity models: $ R\longrightarrow R+f(R)$}.

All these  models look like $\Lambda$CDM, but with an effective non constant 
cosmological term.
\section{Modified Gravity as models for Dark Energy}
The  $\La$CDM model is  the simplest possibility
but, it is worth investigating   more  general modifications, 
possible motivations run from quantum corrections to string models: 
(for a recent review see \cite{NO} and references therein). We shall consider
the modification of the kind $ F(R)=R+f(R)$. Models of this kind are not new and they have been used in the past by many authors, for example as  models for inflation,  $f(R)=aR^2$ \cite{Star80}. Recently their interest in cosmology was triggered by the model  $f(R)=-\mu^4/R$, proposed in order  to describe the current acceleration of the observable universe \cite{capo}.

It is important to stress that  these $F(R)$ models  are conformally 
equivalent to Einstein's gravity, coupled with a self-interacting 
scalar field, Einstein frame formulation. We will consider only the Jordan frame, in which the dynamics of  gravity is described by $F(R)$ with minimally coupled matter. Observations are typically interpreted in this Jordan frame. 

Finally, we would like to  mention the so called viable $F(R)$  models, which  have recently been proposed \cite{Hu}, with the aim to describe the current acceleration with a  suitable choice of $F(R)=R+f(R)$, but also to be compatible  with local stringent gravitational tests of Einstein gravity $F(R)=R$. The main idea is the so called  disappearing of cosmological constant for low curvature, and mimicking the $\Lambda$CDM model for high curvature. Thus, the requirements are:

{\bf a. $f(R) \to 0\,, \quad R \to 0\,$, compatibility with local tests}.  

{\bf b. $f(R) \to  -2 \Lambda_0\,, \quad R \to +\infty$, description of current acceleration}.

{\bf c. Local stability of the matter}.

As a illustration, we recall a recent example of viable model \cite{seba}
$$
f(R)=-\alpha \left( \tanh \left(\frac{b\left(R-R_0\right)}{2}\right) + \tanh \left(\frac{b R_0}{2}\right)\right)
$$
where $R_0$, and $\Lambda_0$  are suitable constants.
Its advantages are a better formulation in the Einstein frame and  a 
generalization that may also include the inflation era.

\section{The de Sitter stability issue}
The stability of the de Sitter solution, relevant for Dark energy, 
 may be investigated in these $F(R)$ models in several ways. 
We limit ourselves to the following  three approaches:

{\bf i. Perturbation of Esq. of motion  in the Jordan frame}.

{\bf ii. One-loop gravity calculation around de Sitter background}.

{\bf iii. Dynamical system approach in FRW space-time}. 

We shall briefly discuss   these  three  approaches. We may anticipate that 
the third one  can easily be extended to more general modified gravitational 
models.
\subsection{Stability  of $F(R)$ model in the Jordan frame} 
The starting point is the trace of the equations of motion, which is 
trivial in
Einstein gravity $R=-\kappa^2 T$, but, for a  general $F(R)$ model, reads
$$
3\nabla^2 f'(R)-2f(R)+Rf'(R)-R=\kappa^2 T\,.
$$
The new non trivial extra degree of freedom  is the Scalaron : $1+f'(R)=e^{-\chi}$. 
Requiring $R=R_0=CST$, one has de Sitter existence condition in vacuum  
$$
R_0+2f(R_0)-R_0f'(R_0)=0\, .
$$
Perturbing around  dS:  $R=R_0+\delta R$, with
$\delta R = - \frac{1+f'(R_0)}{f''(R_0)} \delta\chi$, one arrives at
 Scalaron perturbation Eq.
$$
\nabla^2 \delta\chi - M^2 \delta\chi = - \frac{\kappa^2}{6(1+f'(R_0))}T\, .
$$
One may read off the Scalaron effective mass
$$
M^2\equiv \frac{1}{3}\left(\frac{1+f'(R_0)}{f''(R_0)} - R_0\right)\, .
$$
Thus, if  $M^2>0$, one has  stability of the dS solution and the related 
 condition reads
$$
\frac{1+f'(R_0)}{R_0f''(R_0)}>1\, .
$$
If $M^2<0$,  there is a tachyon and  instability.
Furthermore, one may show that $M^2$ has to be very large in order to 
pass both the local and the astronomical tests and $1+f'(R)>0$, 
in order  to have a positive effective Newton constant. The same result has been obtained within a different more general perturbation approach 
in \cite{fara}.  

\subsection{One-loop $F(R)$ quantum gravity partition function}
 Here we present the generalization to the modified gravitational case  of the study of Fradkin and Tseytlin \cite{fra}, concerning Einstein gravity on dS 
space. One works in the Euclidean path integral formulation, with dS existence condition $2 F_0= R_0  F'_0 $, assumed to be satisfied. The small fluctuations around this dS instanton may be written as
$$
g_{ij}= g_{(0)ij}+h_{ij}\:,\hs
g^{ij}= g_{(0)}^{ij}-h^{ij}+h^{ik}h^j_k+{\cal O}(h^3)\:,\hs
h= g_{(0)}^{ij}h_{ij}\,.\nn 
$$
Making use of  the standard expansion of the tensor field $h_{ij}$ in irreducible components, and making an expansion up to second order in all the fields, 
one arrives at a very complicated Lagrangian density $ L_2 $, not reported here, describing Gaussian fluctuations around dS space. As usual, in order to quantise the model described by $ L_2$,  one has to add gauge fixing and ghost contributions. Then, the computation of Euclidean one-loop partition function reduces 
to the computations of functional determinants. These functional determinants 
are divergent and may be  regularized by the well known zeta-function regularization. The evaluation requires a complicated  calculation \cite{cogno05} and,
neglecting the so called multiplicative anomaly, potentially present in zeta-function regularized determinants (see \cite{vanzo}), one arrives at the  one-loop effective action, here written in the Landau gauge 
\beq
\Ga_{on-shell}&=&\frac{24\pi F_0}{G R_0^2}
+ \frac12\,\log\det\ \aq \ell^2
\at-\lap_2+\frac{R_0}6\ct\cq \nn\\&&
      -\frac12\,\log\det\aq \ell^2 \at-\lap_1-\frac{ R_0}4\ct \cq
\nn\\&&
    +\frac12\,
\log\det \aq \ell^2 \at
-\lap_0-\frac{ R_0}3+\frac{2 F_0}{3 R_0 F_0''}\ct\cq\, \nn . 
\eeq
The last term is absent in the Einstein theory.
As a result,  in the scalar sector, one has  an effective mass 
$M^2=\frac{1}{3}\at \frac{2  F_0}{ R_0  F_0''}- R_0 \ct$. 
Stability requires $M^2 >0$, in agreement with the previous Scalaron analysis,
and with the inhomogeneous perturbation analysis \cite{fara}. 

\subsection{ Dynamical system approach}
This approach has been used by many authors 
\cite{gu,Star80,ellis,amendola,dunsby}. One works in a cosmological setting, namely with a  FRW metric, and the main 
idea consists in rewriting  the generalized Einstein-Friedman equations in an 
equivalent system of first order differential equations, introducing
 new dynamical variables $\Omega_i$
$$
\frac{d}{d t} {\vec \Omega(t)}=\vec v(\vec \Omega(t))\,.
$$
Here the evolution parameter has been denoted by $t$. 
The critical (or fixed ) points are defined by $ \vec v(\vec \Omega_0)=0$.
The key point is: 

{\bf Hartman-Grobman theorem}: 
{\it The orbit structure of a dynamical system  in the neighbourhood of a 
hyperbolic fixed point is topologically equivalent to the orbit 
structure of  the associated linearized dynamical system,
 defined by a stability matrix $M_0$}.

Recall that a hyperbolic fixed point is such that its stability 
matrix $M_0$ does not have vanishing  eigenvalues. In other words the theorem 
states that the flux of a dynamical system in a neighbourhood of a hyperbolic 
fixed point can be continuously deformed to the flux of the related 
linearization. As a result, in order to study the stability of the above 
non linear system of differential Eqs. at critical points, it is sufficient 
to investigate the  related linear system of differential Eqs.:
$$
\frac{d}{d t} \de {\vec \Omega(t)} =M_0 \de \vec \Omega(t)\,,
\quad   M_0 \quad \mbox{Jacobian matrix} 
 \quad \mbox{evaluated at } \vec \Omega_0
$$
The solution of the linearization is well known and the evolution is determined by the signs of the eigenvalues of $ M_0$. As a result, the 
non linear system is stable if all eigenvalues of 
the matrix $M_0$ have negative real parts.

Let us apply this method to study the  stability for $F(R)$ models.
Introducing  new variables are defined by 
$$
\Omega_R=\frac{R}{6H^2}\,,\quad
\Omega_F=-\frac{f(R)-Rf'(R)}{6H^2(1+f'(R))}\,,\quad
\Omega_\rho=\frac{\chi \rho}{3H^2(1+f'(R))}\,,
$$
the dynamical system equivalent  to Einstein-Friedman Eqs. reads
\beq
\frac{d}{d \alpha}\Omega_{R}& =& 2\Omega_R(2-\Omega_R)\Omega_R -
\beta\,\, (1-\Omega_F-\Omega_\rho)\,, \nn \\
\frac{d}{d \alpha}\Omega_{F}& =& 2\Omega_F(2-\Omega_R)+(\Omega_F-\Omega_R)
(1-\Omega_F-\Omega_\rho)\,, \nn \\
\frac{d}{d \alpha}\Omega_{\rho}&=& 
[2(2-\Omega_R)-3(w+1)+1-\Omega_F-\Omega_\rho]
\Omega_\rho\,, \nn
\eeq
here the evolution parameter is $\alpha(t)= \ln a(t)$ and  $w=\frac{p}{\rho}$,
and the function $\beta$ is $\beta(R)=\frac{1+f'(R)}{Rf''(R)}\,$.
Note that one has a complete autonomous system as soon as the quantity $\beta$ can be expressed as a function of $\Omega_i$. This requires the inversion of
$\frac{Rf'(R)-f(R)}{R(1+f'(R)}=\frac{\Omega_F}{\Omega_i} $.
After this inversion, in principle, one has 
$\beta=\beta(\Omega_R,\Omega_F)$, and may close the above system.
The possible problems are: non unique inversions, non trivial domains with
divergent points, ECT.
 The  non linear  algebraic system for critical points is
\beq
0 &=& 2\Omega_R(2-\Omega_R)\Omega_R)-\beta (1-\Omega_F-\Omega_\rho)\,,\nn \\
0 &=& 2\Omega_F(2-\Omega_R)+(\Omega_F-\Omega_R)(1-\Omega_F-\Omega_\rho)\,\nn \\
0 &=& [2(2-\Omega_R)-3(w+1)+1-\Omega_F-\Omega_\rho]\Omega_\rho\,. \nn
\eeq
In vacuum $\rho=0$, namely $ \Omega_\rho=0$, and  de Sitter critical point 
existence condition follows from the solution  
$ \Omega_R=2\,, \Omega_F=1$, namely $ R_0=12H_0\,$ and 
$ R_0=R_0f'(R_0)-2f(R_0)$, in agreement with the other methods. 
In order to investigate the stability of this dS critical point $(2,1,0)$ the associated linear system is
\beq
\frac{d}{d \alpha}\de \Omega_{R}& =& -4\, \de \Omega_R+
2\beta_0\, \de \Omega_F+2 \beta_0 \, \de \Omega_\rho\,,\nn \\
\frac{d}{d \alpha} \de \Omega_{F}& =& -2\, \de \Omega_R+ \de \Omega_F
+\de \Omega_\rho\,, \nn \\
\frac{d}{d \alpha} \de \Omega_{\rho} 
&=& 0\,\, \de \Omega_R+0\,\, \de \Omega_F-3 \gamma\, \de \Omega_\rho\,,\nn 
\eeq
and one can read off  the stability matrix $M_0$, whose eigenvalues are
$ \lambda_1= -3\gamma\,,\quad \gamma >0$ and
$$
\lambda_{2,3}=\frac{1}{2}\at -3\pm \sqrt{25-16 \beta_0 } \ct
$$
The stability condition associated with the de Sitter critical point
requires that the real part of all eigenvalues has to be negative, thus
$$
\frac{1+f'(R_0)}{R_0f''(R_0)}> 1\,,
$$
again in agreement with Scalaron perturbation analysis and one-loop de Sitter 
calculation. In the matter-radiation sector, where $\Omega_\rho$ is non vanishing, other critical points, in general, exist, but their analytical determination, in realistic cases, is problematic, since one has to know explicitly $\beta$ in order to close the system, and  numerical analysis, in general, is necessary.

We conclude recalling that, within this approach, it is not difficult to deal 
with  generalizations of the kind $F(R) \longrightarrow F(R,G,Q,..)\,,$
\cite{carroll05}  where $F$ depends on  arbitrary invariants of tensor 
curvature as $G$, Gauss-Bonnet invariant, $Q$ the square of Riemann tensor, and so on. In the case $F(R,G,Q)$, the associated de Sitter existence solution reads $F=\frac{RF'_{R}}{2}-\frac{R^2}{6}(F'_G+F'_Q)$ and the related  stability condition is (see \cite{monica} and references therein) 
$$
\frac{F'_R+\frac23RF'_Q}{R\aq F''_{RR}+\frac23F'_Q
     +\frac23R(F''_{GR}+F''_{RQ})
      +\frac19R^2(F''_{GG}+2F''_{GQ}+F''_{QQ})
\cq}>1
$$
\section{Concluding remarks}  
Modified gravity may be seen as the phenomenological description 
of a fundamental unknown theory. From this point of view, corrections to 
Einstein-Hilbert action depending on higher order  curvature invariants are 
likely to be expected (Lovelock gravity is an example). 

Among many existing approaches, three methods have been illustrated in 
order to investigate the stability of
these models around de Sitter critical points, and  the dS stability 
conditions has been derived in all the three approaches.  

These methods have owns advantages and problems, but, in our opinion, 
the third one, the dynamical system approach,  permits to study
critical points and stability for modified gravitational 
models depending on arbitrary geometric invariants, 
generalising the results obtained for $F(R)$ models.


\begin{theacknowledgments}
Thanks to G. Cognola, E. Elizalde, S. Nojiri, S. D. Odintsov and L. Vanzo 
useful for discussions.

\end{theacknowledgments}


\begin{thebibliography}{9}

\bibitem{d}
 G.~R.~Dvali, G.~Gabadadze and M.~Porrati,
 \emph{Phys.  Lett.  B }\textbf{485}, 208-214 (2000).


\bibitem{NO}
E.~Copeland, M.~Sami and S.~Tsujikawa, 
\emph{Int. J. Mod. Phys. D} \textbf{15} 1753-1936 (2006); 
S.~Nojiri, S.~D.~Odintsov,
\emph{Int. J. Geom. Meth. Mod. Phys. } \textbf{4}, 115-146  (2007).


\bibitem{Star80}
A.~A.~Starobinsky,
\emph{ Phys. Lett. B } \textbf{91}, 99-146  (1980).

\bibitem{capo}
S.~Capozziello, S.~Carloni and A.~Troisi,
\emph{ Recent Res. Dev. Astron. Astrophys.}  \textbf{ 1}, 625- (2003);  
S.~M.~Carroll, V.~Duvvuri, M.~Trodden and M.~S.~Turner,
 \emph{ Phys. Rev.  D} \textbf{ 70}, 043528 (2004).


\bibitem{Hu}
  I.~Sawicki and W.~Hu,
  \emph{Phys. Rev.  D} \textbf{76},064004 (2007);
  A.~A.~Starobinsky,
  \emph{JETP Lett.}  \textbf{86}, 157-163 (2007);
   S.~Nojiri and S.~D.~Odintsov,
 \emph {Phys. Lett.  B} \textbf{657}, 238-245 (2007). 

\bibitem{seba}
G.~Cognola, E.~Elizalde, S.~Nojiri, S.~D.~Odintsov, L.~Sebastiani and S.~Zerbini,\emph{Phys. Rev.  D} \textbf{ 77}, 046009 (2008).

\bibitem{fara}
  V.~Faraoni,
  \emph{Phys. Rev.  D }\textbf{72}, 124005 (2005).


\bibitem{fra}
 E.~S.~Fradkin, and A.~A.~Tseytlin,
  \emph{Nucl. Phys. B }\textbf{234}, 472-508 (1984).


\bibitem{cogno05}
  G.~Cognola, E.~Elizalde, S.~Nojiri, S.~D.~Odintsov and S.~Zerbini,
  \emph{JCAP }\textbf {0502}, 010 (2005);G.~Cognola and S.~Zerbini,
  \emph{J. Phys. A}  \textbf{39}, 6245-6251 (2006).


\bibitem{vanzo}
E.~Elizalde, L.~Vanzo and S.~Zerbini,
\emph{Comm. Math. Phys. } \textbf{194 }, 613-630 (1998).


\bibitem{gu}
V.~T.~Gurovich and A.~A.~Starobinsky,
\emph{Sov. Phys. JEPT } \textbf{50 }, 844-852 (1979).



\bibitem{ellis}
G.~E.~R.~Ellis and J.~Wainwright Editors, 
\emph{Dynamical System in Cosmology }, Cambridge University Press, Cambridge 
(1997).

\bibitem{amendola}
  L.~Amendola, R.~Gannouji, D.~Polarski and S.~Tsujikawa,
  \emph{ Phys. Rev.  D} \textbf{ 75}, 083504 (2007).

\bibitem{dunsby}
S.~Carloni, P.~K.~S.~Dunsby, S.~Capozziello and A.~Troisi,
\emph{ Class. Quant. Grav.}  \textbf{22}, 4839-4868 (2005);
S.~Carloni, A.~Troisi and P.~K.~S.~Dunsby,
\emph{Some remarks on the dynamical systems approach to fourth order gravity},
  arXiv:0706.0452 [gr-qc].

\bibitem{carroll05}
  S.~M.~Carroll, A.~De Felice, V.~Duvvuri, D.~A.~Easson, M.~Trodden and 
M.~S.~Turner,
 \emph{Phys. Rev.  D} \textbf{ 71}, 063513 (2005).

\bibitem{monica}
  G.~Cognola, M.~Gastaldi and S.~Zerbini,
 \emph{Int. J. Theor. Phys.}  \textbf {47}, 898-910 (2008);
G.~Cognola and S.~Zerbini,
  \emph{Homogeneous cosmologies in generalized modified gravity},
  arXiv:0802.3967 [hep-th].


\end{thebibliography}
\end{document}